# Ontologies for Representing Relations among Political Agents

Carlos Laufer[1][+552135271500], Daniel Schwabe[1][+552135271500], Antonio Busson[1][+552135271500]

[1] Dept. of Informatics, PUC-Rio
R. M. de S. Vicente, 225, Rio de Janeiro, RJ 22453-900, Brazil
`laufer@globo.com`, `{dschwabe,busson}@inf.puc-rio.br`

**Abstract.** The Internet and the Web are now an integral part of the way most modern societies, and corresponding political systems, work. We regard Political systems as the formal and informal political processes by which decisions are made concerning the use, production and distribution of resources in any given society. Our focus in on the sets of agents – Persons and Organizations -that govern a society, and their relations.

We present a set of ontologies aimed at characterizing different kinds of direct and indirect relations that occur within a Political System. The goal is to provide a more semantically precise basis for determining more abstract notions such as "influence".

These ontologies are being used for the Se Liga na Politica project, whose goal is to provide an open linked data database of Political Agents in Brazil. Whereas they are being used in a particular political system, these ontologies can be applied to different political systems.

## 1       Introduction

Since very early times, men have organized themselves to form societies [9], naturally leading to the formation of Political Systems, defined as formal and informal processes by which decisions are made concerning the use, production and distribution of resources in any given society. Societies include Agents – Persons and Organizations – that participate in the processes of its Political System. In doing so, they are driven, and constrained, by the various types of relations that exist among them.

With the advent of the Information Society and the Network Society [1, 2], accelerated by the widespread adoption of the Internet and the Web, information has become a vital resource, both affording and affecting the functioning of Political Systems.

Transparency, the quality that allows participants of the society to know what are the particular processes and agents that are being used in its functioning, is generally regarded as a means to enable checks and balances within Political Systems to prevent misuse by any of the parties involved [6, 4]. One of the forms to increase transparency within a Political System is to provide information about its participants and their relations, as a way to provide additional context.

We have been involved with an initiative to publish a database about Political Agents in Brazil in the form of Linked Data, named "Se Liga na Politica"[1] (SLNP). The data in this database is obtained from several sources, in both automated and non-automated ways. Most of the automated extraction is made from official sources, such as the open data published by the House of Deputies and by the Senate. In addition to such sources, data may also be contributed by individuals, in wiki fashion. In an ideal Linked Data World, such a database would not really be necessary, as institutions responsible for each type of information would publish them in Linked Data form, creating a large Knowledge Graph. In practice, however, we are far from this. The focus of the SLNP project is to establish the *links* between these various "domains of knowledge", taking care not to replicate all of the information published by each source (thereby, somehow, replacing it), focusing on characterizing the *relations* between agents and often omitting other properties that may be of interest of specific communities.

Given the multiplicity of sources, and the nature of the subject matter, this database is designed so that facts are seen as claims made by some agent, and therefore provenance information becomes a "first class citizen" of the domain. One of the main usages for this database is to provide context information for news stories, to allow readers to establish trust in the claimed facts based on their own criteria. In [5] we define more precisely what is the underlying trust process that occurs when consuming online information, and show how provenance information, coupled with the adoption of the nano-publication format, can be used to better support the trust process.

In this paper we present the POLARE ontologies used in the SLNP project, discussing the rationale and modeling choices and the intended uses.

The remainder of this organized as follows. Section 2 presents POLARE; Section 3 discusses related work, and Section 4 draws some conclusions and points to future work.

## 2    POLARE – A Political Agents Relationship Ontology

Before delving in the details of POLARE, we briefly discuss some of the requirements for the ontology and the rationale for the design approach.

### 2.1    Ontologies vs vocabularies

As a general rule, we attempted to used well-known ontologies, such as FOAF[2], ORG[3], SKOS[4], Schema.org[5], etc… as controlled vocabularies to describe concepts in their respective domains. Precisely because these ontologies are very general, they allow many possible uses within other ontologies.

POLARE, in many situations, defines particular ways in which these vocabularies can be used for its purposes; whenever our intended use was incompatible with these ontologies, we used our own vocabulary. In addition, our vocabulary also includes terms to describe concepts not found in any of the better known controlled vocabularies.

---

[1] The expression "Se liga" in Portuguese has a colloquial meaning of "be aware", "pay attention to", as well as "connect yourself". In Portuguese, it reads both as "pay attention to Politics" and "Connect yourself to Politics". A third (indirect) meaning is the reference to Linked Data.

[2] http://xmlns.com/foaf/spec/

[3] https://www.w3.org/TR/vocab-org/

[4] http://www.w3.org/TR/skos-reference

[5] http://schema.org/docs/full.html

POLARE is meant to be used to characterize data in a Linked Data database. We envision that this data may be used in my different ways, for various purposes. To allow such latitude, we have deliberately designed it in a "lightweight" fashion, with few specific inference rules. We believe it is possible to extend it with a more "heavyweight" ontology by including inference rules to further constrain the possible interpretations, for use in specific situations.

One should also remember that POLARE describes statements which are part of nano-publications, and therefore are understood as claims being made by some agent. Therefore, additional care must be taken when including inference rules, as they may be expressing restrictions according do some particular point of view, not necessarily accepted or agreed upon by all users.

The POLARE ontology includes many DataType properties, but since they are not so relevant for characterizing relations, we do not discuss them in this paper.

### 2.2 OWL vs SKOS

OWL is a knowledge representation language, designed to formulate, exchange and reason with knowledge about a domain of interest. OWL can be reasoned with by computer programs either to verify the consistency of that knowledge or to make implicit knowledge explicit [3].

A SKOS concept can be viewed as an idea or notion; a unit of thought. However, what constitutes a unit of thought is subjective, and this definition is meant to be suggestive, rather than restrictive [7].

Our approach is based on a hybrid set of OWL vocabularies and SKOS Concept Schemes, that compose concepts in our domain. OWL is used to define more formal structures where the inference rules can be used to make implicit knowledge. SKOS is used to define concepts that are mainly used for retrieval and navigation tasks, and for which there are many alternative possible schemes.

We have identified several SKOS Concept Schemes that should complement the OWL classes defined in our set of OWL vocabularies. They are used as classifications for specific classes. The rational for choosing to use SKOS as opposed to OWL was based on the generality vs specificity of the concept involved – whenever the concept could be represented in many different ways depending on the particular Political System, we opted to use SKOS. For example, the "classification" of an Organization can be made in many different taxonomies, often non-mutually exclusive – for example, according to fiscal status, legal status, type of ownership/control, etc… We will highlight such uses throughout the description of POLARE.

### 2.3 People and Organizations

The central concepts in POLARE are Persons and Organizations, as they are the Political Agents within a Political System. Given that our goal is to characterize the various kinds of relations between them, we first look at direct relations between Persons, and then look at relations between Persons and Organizations, which establish indirect relations between Persons. We have chosen the FOAF vocabulary to describe Persons, and the ORG vocabulary to describe Organizations, adding relations in the POLARE ontology as needed[6].

The first kind of relations between persons are direct family relations., which are modeled in POLARE as Direct Relationships, shown in **Error! Reference source not found.**. Rather than simply using an owl:ObjectProperty, we chose to model it via reification, as we need to qualify this relation with temporal information. The *directRelProp* property allows specifying what is the family relation; its value, rather than being an rdf:Property, is a skos:Concept, whose value will be taken from a suitable Skos:ConceptScheme. This allows inclusion of certain relations that may not be "formally" accepted as a family relation but may be of interest for some types of analyses, e.g., "co-habitates".

---

[6] For readability purposes, we do not add a prefix to terms of POLARE itself (e.g., pol:hasPost).
Similarly, when it is clear which ontology a term is from (e.g. foaf:Person), we omit the prefix.

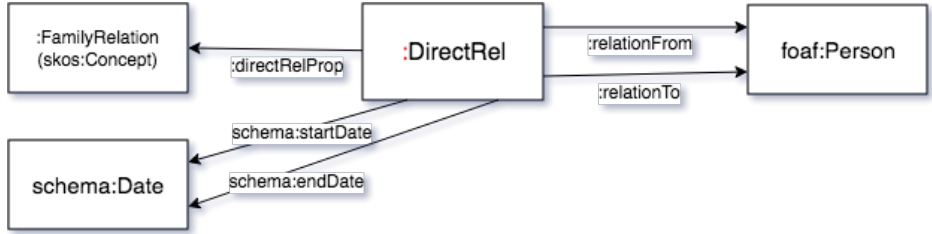

**Figure 1** – Direct relationships between Persons in POLARE

Direct relations between Organizations area already contemplated in the ORG vocabulary (e.g., org:hasSubOrganization, org:HasUnit). In the future, we may extend relations between Organizations in an analogous way as the *DirectRel* approach, to represent more nuanced relations such as shareholding control.

Persons are related to Organizations chiefly through occupying positions within the Organizations, modeled using the org:Post class[7]. The main terms of the POLARE ontology come from W3C's ORG Ontology and are shown in **Error! Reference source not found.**. It is important to notice the reification of the relation "occupies" between Person and Post via the Membership class, which is fundamental to allow us to represent properties of this relation, as will be discussed later.

The property org:role is used to identify an org:Post occupied by a person in an org:Organization.; its value is an org:Role that is a skos:Concept defned in a specific SKOS Concept Scheme.

We have added the *hasPost* relation linking org:Membership with org:Post because we need to represent that a Post might be occupied by different persons in different periods of time.

Both Post and Membership may have start and end dates associated to them. The dates associated to a Post refer to a time period when the Post exists, for example, for a post in the House of Representatives, it corresponds to a Legislature, which defines the mandate of the elected person. The Membership dates refer to the period in which the Person actually occupies the Post, as it is possible for a Person to temporarily leave the Post for a short period of time within the mandate.

An important characteristic of most public organizations that make up a Government is that they have fixed number of Posts. For example, the House of Deputies has a fixed number of seats (Posts), and a particular seat can be occupied by at most one person for any given moment. For this reason, although the ORG ontology allows direct org:memberOf relations between Agents and Organizations, we have chosen not to use this relation, requiring always an org:Post to exist to mediate the relation between Persons and Organizations.

---

[7] Notice that "Post" here refers to a "Position", not to a blog post.

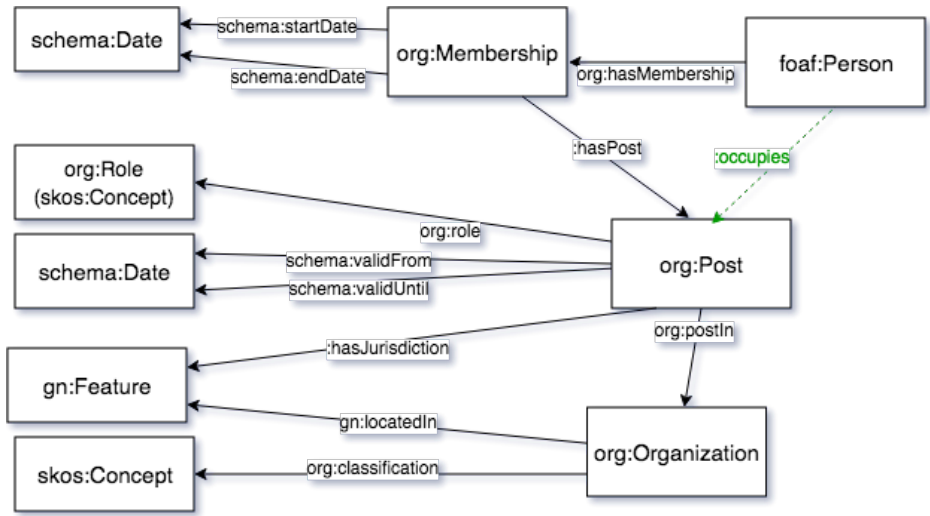

**Figure 2-** Persons and Organizations the POLARE ontology.
Inferred relations are shown as dotted lines.

In Political Systems, and also for many Organizations, it is important to know if some Post was filled through a *Referral*, i.e., that some foaf:Agent indicated (nominated) a foaf:Person to occupy an org:Post. POLARE can represent which Agent has referred some Person to occupy a Post in an Organization, shown in **Error! Reference source not found.**. We also reified the *refers* property using a *Referral* class, to allow including properties related to the Referral itself.

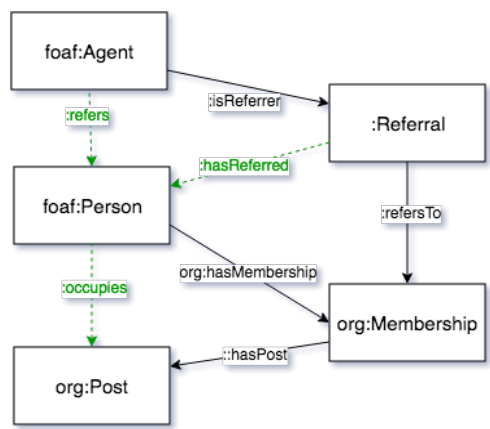

**Figure 3 –** Modeling Referrals in POLARE

Several indirect relationships of interest between Persons can be captured simply by the fact that these Persons occupy Posts in the same Organization. For example, the fact that two congressmen belong to the same party (which is an Organization), or that they were once colleagues in a company in the private sector or in a department in the executive branch.

## 2.4 Legislative aspects

One of main activities in a Political System is the elaboration and enactment of legislation that establish and regulate many of its processes. Figure 4 shows a diagram of the part of POLARE showing relations involved in these activities.

Legislators (foaf:Persons) make (dc:creator) *Propositions* (known by several terms in particular Political Systems, e.g., "Bill" in the US and Uk, "Projeto de Lei" in Brazil, "Projet/Proposition de Loi" in France) that undergo a voting process to become a *Law*. The voting process takes place in *Sessions* (which are schema:Event) comprised to a series of *VoteEvent*s. Each *VoteEvent* requires legislators (*Voter*) to issue a *Vote* regarding some *Disposition* (e.g., substitution, amendment, approval, etc…) relative to the *Proposition* in question. The actual vote (e.g., "yes", "no", "abstain", etc…) is modeled as a skos:Concept that is the value of the *vote* property for instances of class *Vote*.

The class *Voter* stands for the role of the Person in voting and is necessary because it is important to preserve the relation with the Party (which is an Organization) to which the legislator belongs. Since in many Political Systems legislators can change their party affiliations throughout their mandate, it is important to record the particular affiliation in effect at the time of the vote. It is true that this information could be inferred by retrieving the set instances of *Membership* between the Person and Political Parties (as instances of *Organizations*), checking each for its associated time interval (*startDate* and *endDate*), and determining the one that was in effect at the *startDate* of the *VoteEvent*. Nevertheless, we decided to include this information directly in the recorded data, since it is directly provided by the datasources we used in our project and is also commonly provided by datasources reporting legislative actitivy worldwide.

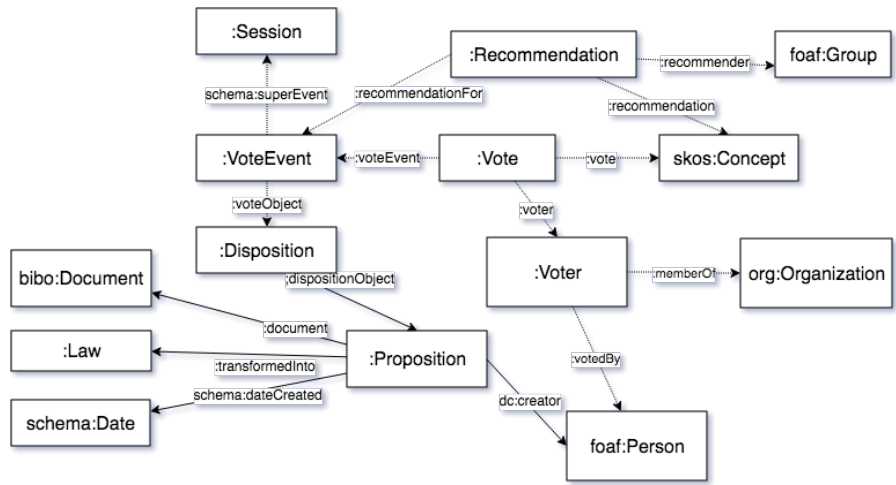

**Figure 4** –Relations derived from legislative activity in POLARE

For many Political Systems, certain *VoteEvent*s may have a voting Recommendation issued by some group of voters (foaf:Group), for example, a caucus.

## 2.5 Electoral Process

A central aspect of a Political System is the way public officials are chosen, and involvement in this process is another indirect way in which relations between Political Agents are established. Figure 5 shows a portion of POLARE modelling the electoral process.

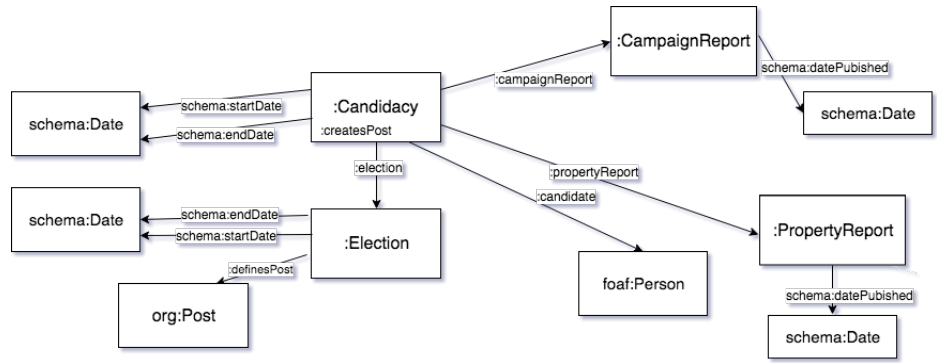

**Figure 5** Electoral activity in POLARE

An *Election* defines a number of org:Posts (e.g. seats in the Senate) for which Persons can run for. In other words, a *Candidacy* models that a foaf:Person is a *candidate* in an *Election* for an org:Post. Notice that although we express this in terms of an election, this pattern could be applied to many other selection processes.

For many Political Systems, transparency requirements include filing a *CampaignReport* to detail the finances (incomes and expenses) incurred during the electoral campaign. Another requirement is the filing of a *PropertyReport* detailing the assets owned by the candidate at the time of the candidacy. It is expected that this report can be compared to a similar one issued when the person leaves the office, enabling verification of possible irregularities.

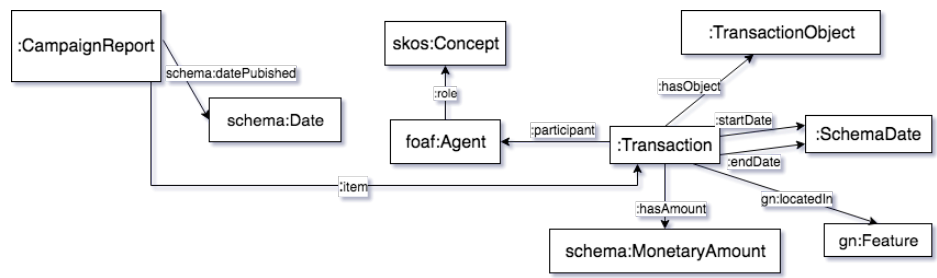

**Figure 6** Transactions in POLARE

The campaign report is a list of *Transaction*s. Each *Transaction* involves a number of foaf:Agents each assuming a *role* (e.g., seller, buyer, guarantor, etc…) which is skos:Concept chosen from a suitable skos:ConceptScheme. The transaction involves an object (TransactionObject, which is either a schema:Product or schema:Service) which is rendered in exchange for a certain amount.

It should be noted that although this model for transactions is presented here within the context of a campaign report, it may be used to record any transactions involving Political Agents, not necessarily during an electoral campaign. Thus, two or more Political Agents are (indirectly) related if they participated in the same *Transaction*.

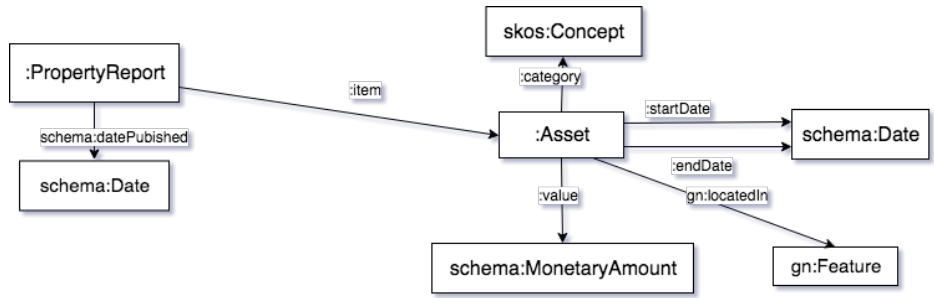

**Figure 7** Asset list in POLARE

As shown in Figure 7, shows the POLARE fragment representing a *PropertyReport,* which is simply a list of *Assets* owned by the candidate in an *Election*. Note that, this list will only characterize relations between Political Agents under the presence of some *Transaction* that establishes how that *Asset* came into possession of it owner. As such, an *Asset* is related to *TransactionObject.*

## 2.6 Legal actions

Another important aspect of Political Systems is the way it handles violations of its established norms. When such a situation occurs, it is dealt with through legal actions. Figure 8 shows the fragment of POLARE representing the relations involved in a *LegalCase*.

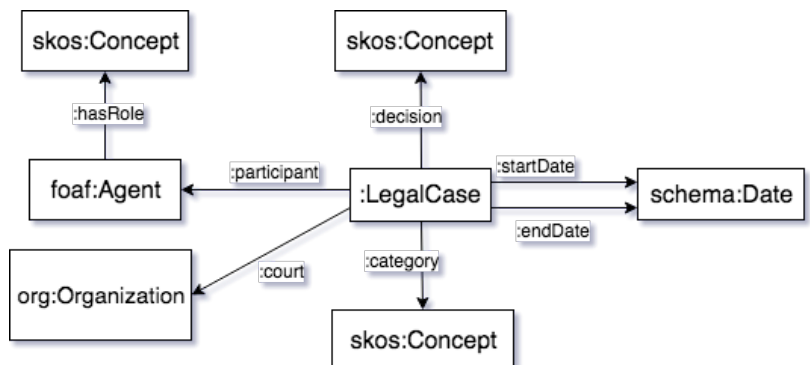

**Figure 8** – POLARE representation of legal cases

Each participant is a foaf:Agent that fulfills a role (hasRole) in the case, such as plaintiff, defendant, judge, attorney, etc…Once more, such roles are modeled as skos:ConceptSchemes, which must exist for the various Political Systems. Thus, two Political Agents are (indirectly) related if they participated, in any role, in the same *LegalCase*.

## 3   Related Work

Two projects have proposed similar ontologies as POLARE - the Poderopedia Project[8] and the The Popolo Project[9] Poderopedia has built a database about Political Agents allowing instantiations for different countries. It is a Linked Data database using the PoderVocabulary[10], which shares similar concepts with POLARE regarding Persons and Organizations, in some cases reaching a finer grain modeling of some relations, notably between Organizations. It took the approach of using OWL as much as possible, so several of the concepts modeled using SKOS in Polare are modeled as OWL classes. The PoderVocabulary does not detail legal actions, elections nor transactions, and does not have the Referral relation.

The Popolo Project is an initiative to define data interchange formats and data models for governments, in the context of Open Government. They define a set of classes that cover, basically, the specification of Persons and Organizations, using Posts and Memberships to relate Persons to Organizations. Popolo also defines a set of classes related to voting processes. The main principle of Popolo is reuse. The classes are defined as a set of properties of well-known vocabularies, including ORG, FOAF, geonames, schema.org, etc. POLARE defines the linking between Persons to Organizations in a similar way. POLARE introduces a set of relations that are not contemplated by Popolo, as family relations, referrals, legal actions or the elections. Popolo does not include provenance information.

## 4   Conclusions and Future Work

We have presented the POLARE ontology, aimed at representing the various types of relations among agents in a Political System. Special attention was given to accommodate various kinds of Political Systems, mostly by the appropriate choice of SKOS Concept Schemes. We have already used it to model relations in the Brazilian Political System in the *Se Liga na Politica* project.

There are several directions in which we are continuing this work. First, we have observed that relations occurring in the academic environment are often relevant, notably advising a Master's dissertation or a PhD thesis, and we will extend POLARE to be able to express them.

As second area of interest is capturing more nuanced relations among Organizations, beyond what can be expressed in the ORG ontology. For example, campaign contributions are often made by several companies that belong to the same "economic group", but the relations that connect companies in the same group go beyond the ones available in ORG. Examples of such relation types are discussed in the OpenCorporates website[11].

As already noted, the modeling of the electoral process could, with some small changes, be generalized to any selection process, although it should remain that such generalizations should only be considered for processes in which new and relevant kinds of relations among political agents would emerge (e.g., analogously to campaign contributions in election processes).

Throughout the text we have mentioned several skos:ConceptSchemes that should be used to classify class instances and relation types. We plan to develop such schemes for various Political Systems.

Although we have presented POLARE as an ontology using several vocabularies, the POLARE vocabulary itself would not be enough to characterize the intended uses as discussed here in an informal way. We are investigating using ShEx[12] or SHACL[13] as a way to describe more precisely the intended graph structures in POLARE.

---

[8] http://www.poderopedia.org

[9] http://www.popoloproject.com/

[10] http://dev.poderopedia.com/documentation/index

[11] https://blog.opencorporates.com/2013/10/16/understanding-corporate-networks-part-1-control-via-equity/

We have also been investigating the idea of reifying the properties using the singleton property approach [8], which allows modeling properties (actually property instances) as individuals and as Properties. Figure 10 shows a snippet of the POLARE schema using this approach, and the corresponding OWL fragment of an instance. We want to investigate if the differences in these approaches could reveal a better way of constructing queries to inspect data in the database.

Another implementation aspect we will investigate is the addition of access control to portions of the database. We envisage that certain communities (e.g., reporters) may want to enter information to the database but restrict, at least for a period of time, general access to it.

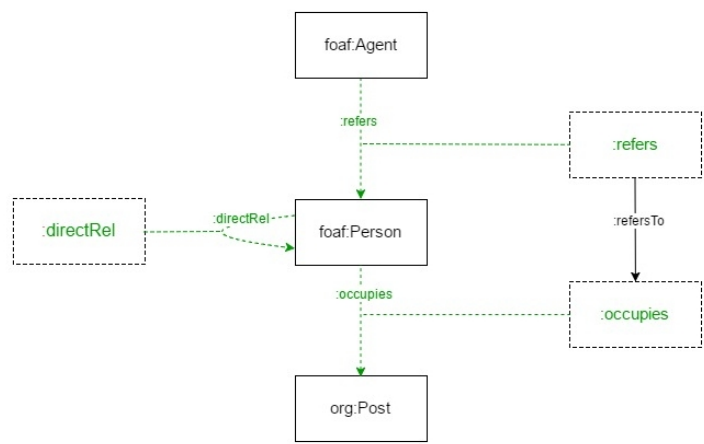

**Figure 9 -** Representing properties using the Singleton approach.

```
:John rdf:type owl:NamedIndividual ,foaf:Person ;
        foaf:name "John Doe"^^xsd:string ;
        :occupies_1 :Post_1 ;
:occupies_1 rdf:type owl:NamedIndividual, owl:ObjectProperty,
            org:Membership ;
            schema:startDate "2015-01-01"^^xsd:date ;
            :singletonPropertyOf :occupies.
:occupies rdf:type owl:NamedIndividual, owl:ObjectProperty.
```

**Figure 10 –** Representing relations as singletons

## 5    References

1. Castells, Manuel (2010). The Rise of the network society (2 ed.). ISBN 9781405196864. Retrieved 1 December 2016.

---

[12] http://shex.io/shex-semantics/
[13] https://www.w3.org/TR/shacl/